\documentclass[sigconf,nonacm]{acmart}
\newcommand{\cmark}{\textcolor{green!50!black}{\checkmark}}
\newcommand{\xmark}{\textcolor{red!70!black}{$\times$}}

\usepackage{booktabs}
\usepackage{amsmath}
\usepackage{tabularx}
\usepackage{xcolor}
\usepackage{colortbl}
\usepackage{array}
\usepackage[most]{tcolorbox}
\usepackage{graphicx}
\usepackage{tikz}

\newcommand{\abstain}{\textsc{Abstain}}

\DeclareMathOperator{\cpc}{GCPC}

\definecolor{ChkBlueBg}{HTML}{EAF1FF}
\definecolor{ChkBlueLine}{HTML}{6F8FD8}
\definecolor{ChkBlueText}{HTML}{315A9B}

\definecolor{ChkPinkBg}{HTML}{FCECF1}
\definecolor{ChkPinkLine}{HTML}{D98CA2}
\definecolor{ChkPinkText}{HTML}{91485D}

\definecolor{ChkGrayBg}{HTML}{F2F3F5}
\definecolor{ChkGrayLine}{HTML}{B7BDC5}
\definecolor{ChkGrayText}{HTML}{555E69}
\definecolor{ChkRule}{HTML}{DDE1E6}

\definecolor{acegreen}{HTML}{E7F1EB}
\definecolor{aceblue}{HTML}{E8EEF5}
\definecolor{aceorange}{HTML}{F6EAE4}
\definecolor{acegray}{HTML}{F1F2F3}
\definecolor{acepos}{HTML}{2F6B4F}
\definecolor{aceneg}{HTML}{98513B}
\definecolor{ACEBlueBand}{HTML}{E8F0FF}
\definecolor{ACEBlueCell}{HTML}{F1F5FF}
\definecolor{ACEBlueText}{HTML}{274D90}

\definecolor{ACEPinkBand}{HTML}{FCECEF}
\definecolor{ACEPinkCell}{HTML}{FFF1F4}
\definecolor{ACEPinkText}{HTML}{8B4154}

\definecolor{ACEGrayBand}{HTML}{F1F3F5}
\definecolor{ACEGrayText}{HTML}{4F5964}

\newtcbox{\assertionroute}{
  on line,
  arc=2pt,
  boxrule=0.45pt,
  boxsep=0.6pt,
  left=2.5pt,
  right=2.5pt,
  top=1pt,
  bottom=1pt,
  colback=ChkBlueBg,
  colframe=ChkBlueLine,
  coltext=ChkBlueText,
  fontupper=\sffamily\bfseries\tiny
}

\newtcbox{\generatedroute}{
  on line,
  arc=2pt,
  boxrule=0.45pt,
  boxsep=0.6pt,
  left=2.5pt,
  right=2.5pt,
  top=1pt,
  bottom=1pt,
  colback=ChkPinkBg,
  colframe=ChkPinkLine,
  coltext=ChkPinkText,
  fontupper=\sffamily\bfseries\tiny
}

\newtcbox{\judgeroute}{
  on line,
  arc=2pt,
  boxrule=0.45pt,
  boxsep=0.6pt,
  left=2.5pt,
  right=2.5pt,
  top=1pt,
  bottom=1pt,
  colback=ChkGrayBg,
  colframe=ChkGrayLine,
  coltext=ChkGrayText,
  fontupper=\sffamily\bfseries\tiny
}

\setcopyright{none}
\acmConference[WSDM '27]{The 20th ACM International Conference on Web Search and Data Mining}{February 15--19, 2027}{TBD}

\begin{document}
\title{Grounded Checklist Partial Credit for Agent Skill Trajectories}

\author{Suliu Qin}
\affiliation{%
  \institution{Xi'an Jiaotong-Liverpool University}
  \city{}
  \country{}
}
\email{Suliu.Qin22@student.xjtlu.edu.cn}

\author{Lu Yin}
\affiliation{%
  \institution{University of Surrey}
  \city{}
  \country{}
}
\email{l.yin@surrey.ac.uk}

\author{Xilu Wang}
\affiliation{%
  \institution{University of Surrey}
  \city{}
  \country{}
}
\email{wangxilu@surrey.ac.uk}

\begin{abstract}
Language-model agents increasingly tackle long-horizon tasks in interactive environments, yet their evaluation commonly relies on task-level success rates by reducing an entire execution trajectory to whether the task passes an official verifier. This binary score hides partial progress and is particularly limited for procedural agent skill evaluations, since a skill can alter execution without changing the final outcome. While checklists provide finer-grained evaluation by scoring individual task requirements, costly manual authoring and unreliable automatic generation make trustworthy evaluation difficult to scale. To address these challenges, we introduce Grounded Checklist Partial Credit (GCPC), a human-governed and LLM-instantiated partial-credit evaluation of agent trajectories. Humans define reusable rules once, from which an LLM instantiates a task-specific checklist grounded in the task instruction and official verifier. To keep judgment tied to evidence, a judge scores each item from execution log evidence alone and abstains when evidence is missing. A separate scripted step then applies the official verifier outcome to the score. Across a 4{,}455-trajectory, deduplicated SkillsBench evaluation population, GCPC better discriminates official PASS and FAIL outcomes than holistic judging on the shared subset (AUC 0.689 vs. 0.619). Human evaluation on 96 trajectories from 12 tasks shows that GCPC aligns more closely with human assessments of progress. Applied to 1{,}946 matched with/without-skill pairs,
GCPC exposes the effects hidden by pass@1: among 879 pairs whose binary outcome
does not change, 20.9\% improve by more than $0.10$ while 18.7\% regress by the
same margin. The GCPC pipeline also transfers to Terminal-Bench and SWE-bench, demonstrating applicability beyond skill-conditioned evaluation.
\end{abstract}

\maketitle
\fancyhead[LE]{}
\fancyhead[RE]{}
\fancyhead[LO]{}
\fancyhead[RO]{}
\fancyfoot[C]{\thepage}

\section{Introduction}

\begin{figure*}[t]
\centering
\includegraphics[width=0.8\linewidth]{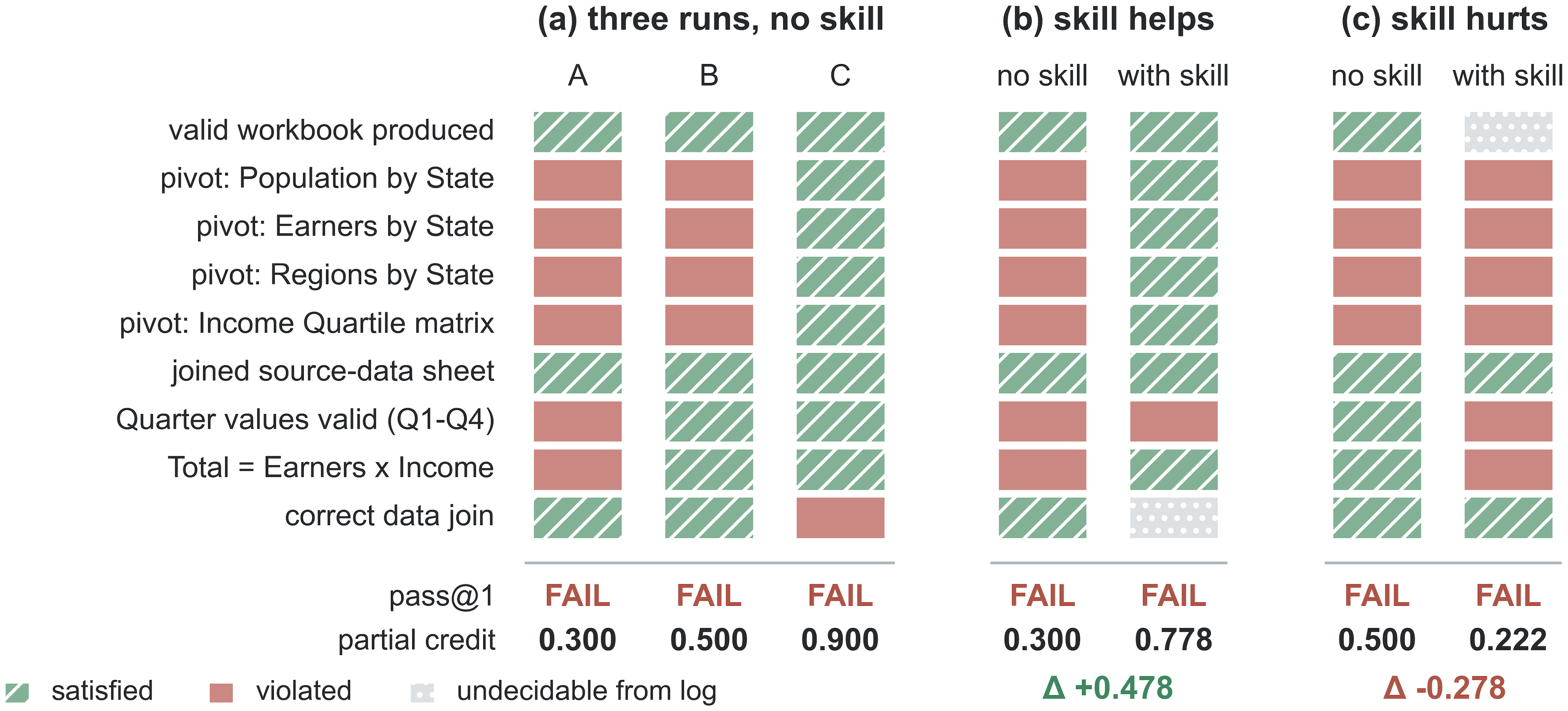}
\caption{
Checklist partial credit reveals execution differences hidden by binary outcomes on the SkillsBench task \texttt{sales-pivot-analysis}.
(a) Three runs all receive an official FAIL outcome and are therefore indistinguishable under pass@1, but satisfy different numbers of checklist requirements, yielding partial-credit scores of $0.300$, $0.500$, and $0.900$. In the last run, the execution log supports all nine checklist items, but the official verifier rejects the saved artifact. The verifier outcome is therefore included as a tenth violated endpoint item, giving $9/10=0.900$.
(b, c) Matched runs with and without skill access remain failed under binary evaluation, while partial credit reveals substantial improvement in one pair ($\Delta=+0.478$) and regression in the other ($\Delta=-0.278$).
}
\label{fig:intro}
\end{figure*}


Recently, large language model (LLM) agents have drawn significant attention and been applied to real-world problem-solving, including software engineering, web browsing, mathematics, and scientific discovery \citep{kapoor2026holistic,mohammadi2025evaluation}. Solving these tasks typically requires a long sequence of interdependent actions, where each step produces intermediate results, such as tool outputs, partial file edits, or retrieved evidence. Despite these achievements, evaluation of LLM agents remains challenging, limiting the reliability required for real-world deployment. Long-horizon tool-using agents are commonly evaluated through task-level success rates: an execution is marked as successful if it satisfies an environment or verifier, and as failed otherwise \citep{ma2024agentboard,yao2025taubench,li2026skillsbench}. This evaluation compresses the entire trajectory into a binary score, which provides limited insights into these intricate processes.  This limitation is particularly severe when evaluating agents with procedural skills, because a skill is designed to change how an agent interacts with the environment, even when the final task outcome remains unchanged. As illustrated in Figure~\ref{fig:intro}, binary task-level evaluation can hide substantial differences in agent execution. Figure~\ref{fig:intro} (a) shows three runs of the same task that all receive a FAIL outcome and therefore the same pass@1 score, despite satisfying three, five, and nine of the nine task requirements, respectively. Figures~\ref{fig:intro}(b) and~\ref{fig:intro}(c) further illustrate this limitation for procedural skills. Introducing a skill leads to more satisfied requirements in one case and fewer in the other, while the task-level outcome remains FAIL in both. These examples show that binary evaluation captures neither partial task progress nor how a skill changes the execution process, motivating a finer-grained evaluation of agent trajectories.

Fine-grained evaluation offers a natural alternative by decomposing a task into
requirements and scoring them separately
\citep{ma2024agentboard,odysseys2026,agentprocessbench2026}. Rather than reporting only the final outcome, it reveals which parts of the task are completed and how the execution changes across runs. The unresolved
problem is how to obtain trustworthy requirements at scale. Human-authored
rubrics can provide precise evaluation criteria, but require repeated task-specific construction and
validation \citep{odysseys2026}. LLM-generated checklists reduce this authoring
cost \citep{cook2024tick,wei2025rocketeval,zhou2026autochecklist}, but may omit important
requirements, duplicate them at inconsistent granularity, or prioritize
details that do not reflect human evaluation standards
\citep{zhang2026rubricbench}. A separate risk arises at scoring time, as an LLM judge may then reward a plausible-looking log even when the required artifact state is absent, inheriting known
holistic and positive biases in process judgment
\citep{agentprocessbench2026,zhang2026rubricbench,zheng2023mtbench,ye2024justice,siro2026learning}. Hence, the central challenge is not decomposition alone, but a \emph{scalability--trustworthiness dilemma}: an evaluation instrument must scale across tasks while keeping every item and every score auditable.

To make fine-grained evaluation scalable without sacrificing trustworthiness, we introduce \emph{Grounded Checklist Partial Credit} (GCPC), a human-governed and
model-instantiated pipeline for effectively synthesizing verified task-specific checklists and scoring agent trajectories. To achieve this, human experts define a small set of reusable rules once, specifying how items should be written, how they must be grounded in the task instruction, how this grounding is automatically validated, and a five-category scheme to audit checklist coverage and diagnose execution failures. An LLM then applies these rules to each task's instruction and official verifier \citep{yao2025taubench,li2026skillsbench} to generate a task-specific checklist, without seeing any trajectory, outcome, or reference solution. A separate judge scores each item in the checklist using evidence from the execution log, and summarizes these scores as the GCPC score, the fraction of decidable requirements that are satisfied. The official verifier outcome is incorporated only after log-based scoring through fixed postprocessing, keeping trajectory evidence separate from information available only to the verifier. The central contribution is an auditable synthesis-and-scoring pipeline for partial credit on checklists, rather than a new aggregation formula. The pipeline allows humans to govern how items are derived, validated, judged, and combined with the official outcome. Our main contributions are summarized below:

\begin{itemize}
\item We introduce \emph{Grounded Checklist Partial Credit} (GCPC)
for fine-grained evaluation of agent skill trajectories beyond binary task outcomes.
Instead of reducing an execution to PASS or FAIL, GCPC measures which task
requirements are satisfied and how much progress an agent makes toward task
completion. The GCPC score therefore captures execution differences that remain invisible under task-level success rates, including changes between runs with the same final outcome.
\item GCPC makes such partial-credit evaluation scalable and auditable through
a human-governed and LLM-instantiated checklist synthesis procedure. Humans define reusable rules for item synthesis, provenance, and validation once, while an LLM applies them to each task's instruction and official verifier. Each item is grounded in the task specification, assigned an explicit verification route, and mechanically validated before scoring.
\item We validate GCPC against holistic trajectory scoring and independent human judgments on SkillsBench, showing that it achieves a higher AUC (0.689 versus 0.619) and closer agreement with human assessments than holistic scoring. Applied to 1{,}946 matched executions with and without procedural skills, GCPC reveals substantial improvements and regressions hidden by unchanged binary outcomes. GCPC further transfers to Terminal-Bench and SWE-bench, demonstrating applicability beyond skill-conditioned evaluation.
\end{itemize}

\section{Related Work}
\label{sec:related}

Our work lies at the intersection of fine-grained agent evaluation,
rubric- and checklist-based judging, and the evaluation of agent skills.
Table~\ref{tab:positioning} positions representative systems using design
properties that can be stated consistently across these research threads.

\begin{table*}[t]
\centering
\small
\renewcommand{\arraystretch}{1.0}
\setlength{\tabcolsep}{3.5pt}

\caption{Positioning against representative structured-evaluation methods
($\cmark$ = explicitly uses the design, not a preference judgment).
\emph{Source quote}: item-level task-text quotation; \emph{verifier link}:
explicit connection to executable benchmark checks.}
\label{tab:positioning}

\begin{tabular*}{\textwidth}{
  @{\extracolsep{\fill}}
  l c c c c c c c c
  @{}
}
\toprule
\textbf{Method}
& \textbf{Auto-built}
& \textbf{Per-task}
& \textbf{Source quote}
& \textbf{Verifier link}
& \textbf{Item verdict}
& \textbf{Abstain}
& \textbf{Partial credit}
& \textbf{Trajectory} \\
\midrule

\multicolumn{9}{l}{\textit{Rubric- and checklist-based evaluation}}\\
TICK \citep{cook2024tick}
& \cmark & \cmark & \xmark & \xmark
& \cmark & \xmark & \cmark & \xmark \\

CheckEval \citep{lee2025checkeval}
& \cmark & \xmark & \xmark & \xmark
& \cmark & \xmark & \cmark & \xmark \\

RocketEval \citep{wei2025rocketeval}
& \cmark & \cmark & \xmark & \xmark
& \cmark & \xmark & \cmark & \xmark \\

AutoChecklist \citep{zhou2026autochecklist}
& \cmark & \cmark & \xmark & \xmark
& \cmark & \xmark & \cmark & \xmark \\

\midrule
\multicolumn{9}{l}{\textit{Fine-grained agent evaluation}}\\
AgentBoard \citep{ma2024agentboard}
& \xmark & \cmark & \xmark & \cmark
& \cmark & \xmark & \cmark & \cmark \\

WebCanvas \citep{pan2024webcanvas}
& \xmark & \cmark & \xmark & \cmark
& \cmark & \xmark & \cmark & \cmark \\

Mind2Web~2 \citep{gou2025mind2web2}
& \cmark & \cmark & \xmark & \xmark
& \cmark & \xmark & \cmark & \xmark \\

Odysseys \citep{odysseys2026}
& \cmark & \cmark & \xmark & \xmark
& \cmark & \xmark & \cmark & \cmark \\

\midrule
\multicolumn{9}{l}{\textit{Agent-skill evaluation}}\\
SkillsBench \citep{li2026skillsbench}
& \xmark & \cmark & \xmark & \cmark
& \xmark & \xmark & \xmark & \cmark \\

SkillJuror \citep{skilljuror2026}
& \xmark & \cmark & \xmark & \cmark
& \xmark & \xmark & \xmark & \cmark \\

\midrule
\textbf{Checklist partial credit (ours)}
& \cmark & \cmark & \cmark & \cmark
& \cmark & \cmark & \cmark & \cmark \\

\bottomrule
\end{tabular*}
\end{table*}

\subsection{Agent Evaluation Beyond Task Success}
\label{sec:rw-agent-evaluation}

Agent benchmarks commonly summarize an execution through final-state success,
which supports reproducible comparison but reveals little about partial
progress. Several benchmarks therefore introduce intermediate evaluation
structures. AgentBoard supplements success rate with manually specified
subgoals and a progress-rate metric \citep{ma2024agentboard}, while WebCanvas
scores key states reached during web interaction
\citep{pan2024webcanvas}. Mind2Web~2 constructs task-specific rubric trees and
judge agents to evaluate long-horizon search results
\citep{gou2025mind2web2}. Odysseys similarly decomposes long-horizon web tasks
into graded rubric checkpoints and reports closer agreement with humans than
holistic trajectory judging \citep{odysseys2026}. These works demonstrate the
value of measuring intermediate accomplishments, but their evaluation
structures are generally created together with the benchmark. We study the
complementary setting in which task instructions, executable verifiers, and
trajectory corpora already exist and a fine-grained instrument must be
constructed retrospectively. Work on process-level judging further shows that
direct judgments of agent steps can be sensitive to label definitions and
judge bias \citep{agentprocessbench2026,lightman2024verify,sharma2026willful}; our items instead describe task
requirements and permit abstention when the saved trajectory does not expose
sufficient evidence.

\subsection{Rubric- and Checklist-Based Evaluation}
\label{sec:rw-checklists}

Rubrics and checklists have emerged as alternatives to a single holistic judgment with decisions over explicit items by decomposing agent trajectories into explicit items and assessing each item separately. Human-authored requirements provide strong semantic control, as in InFoBench, WildBench, and HealthBench
\citep{qin2024infobench,lin2024wildbench,healthbench2025}, but require
substantial task-specific labor. Alternatively, automated approaches derive items
from instructions, candidate responses, quality dimensions, or feedback.
TICK generates instruction-specific yes/no questions
\citep{cook2024tick}; CheckEval decomposes human-selected quality dimensions
into Boolean questions \citep{lee2025checkeval}; and RocketEval uses
instance-specific checklists with lightweight evaluators
\citep{wei2025rocketeval}. AutoChecklist organizes such methods as composable
generator--refiner--scorer pipelines \citep{zhou2026autochecklist}. However, automatically generated items can omit important requirements, vary in granularity, or include items that are difficult to verify
\citep{song2024finesure,saadfalcon2025unittests,que2024hellobench,interacteval2024,dineen2025qalign}.

Decomposition alone does not guarantee a trustworthy instrument. RubricBench
reports a substantial performance gap between model-generated and
human-authored rubrics \citep{zhang2026rubricbench}, while inserting checklists
as soft guidance does not necessarily improve holistic judging
\citep{furuhashi2025checklists,ding2026adarubric,liu2026openrubrics,rlcf2025,wan2026rubricguided}. These findings motivate our emphasis on
item provenance. Our method constrains every generated item with a
verbatim task quotation, links assertion-route items to executable verifier
checks, freezes the checklist before any trajectory is observed, and separates
item judgment from deterministic aggregation and the official-outcome step.

\subsection{Evaluation of Agent Skills}
\label{sec:rw-skills}

Agent skills package procedural knowledge for use at inference time
\citep{wang2023voyager,xu2026skills,jiang2026sok,skillsurvey2026,li2026ecosystem}. SkillsBench evaluates their marginal
utility under conditions without skills, curated skills, and self-generated skills,
with deterministic task verifiers as the primary outcome
\citep{li2026skillsbench,skillsbench2026granularity}. SkillJuror further studies how skill organization
changes resource access and runtime behavior while keeping task knowledge
fixed \citep{skilljuror2026}. These studies show that skills can affect both task outcomes and execution behavior, but their comparative signal remains either the verifier's final result or diagnostics computed over the whole trajectory. As a result, they provide limited insights into partial progress when a skill changes which task requirements are satisfied without changing the final PASS or FAIL outcome. Therefore, we do not introduce another skill benchmark. Instead, we construct a verifier-grounded checklist partial credit score to analyze the trajectories that existing agent-skill benchmarks already produce.


\section{Method: Grounded Checklist Partial Credit}
\label{sec:method}

\subsection{Setting and Pipeline Overview}
\label{sec:setting}

\begin{figure*}[t]
\centering
\includegraphics[width=\textwidth]{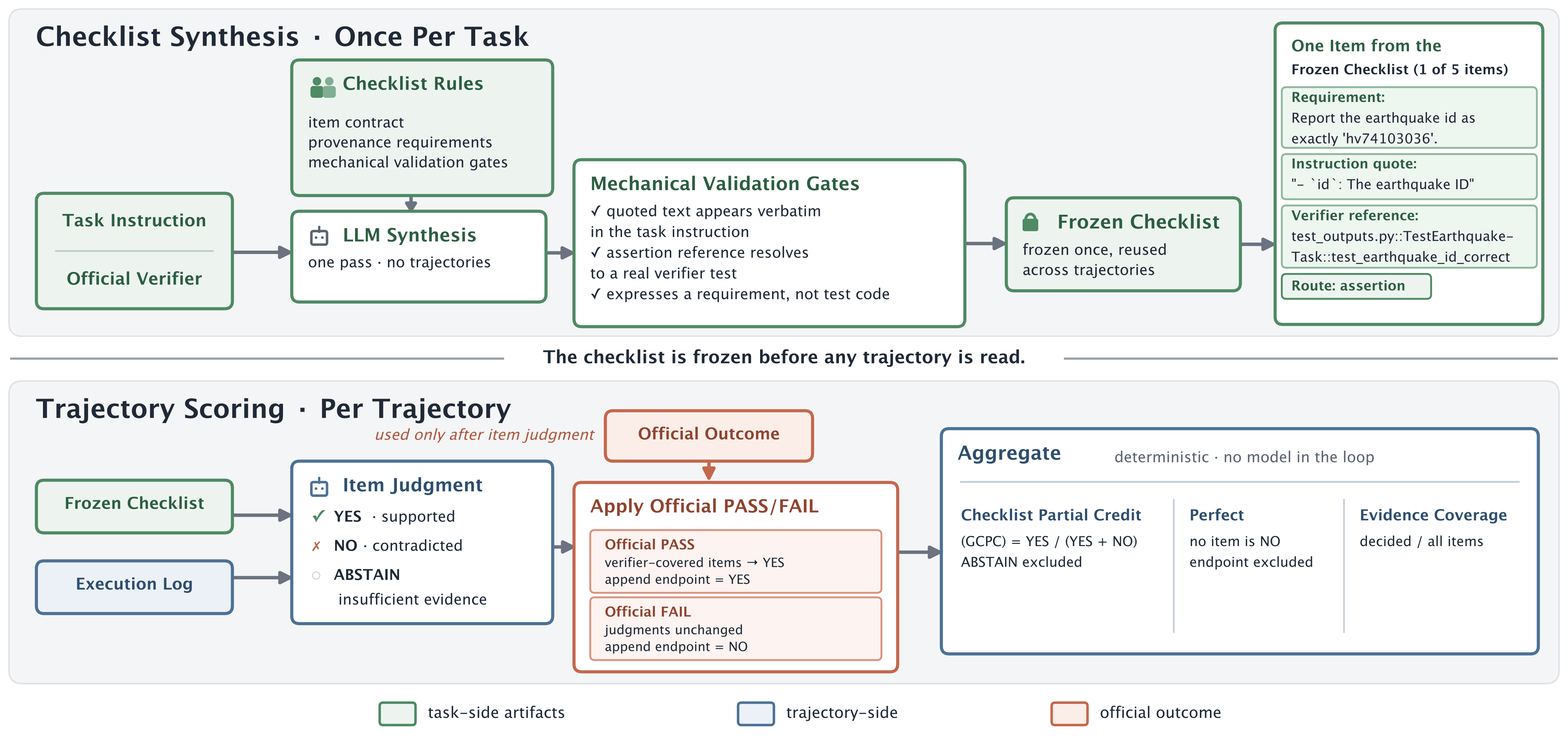}
\caption{Overview of the GCPC pipeline. Checklist synthesis runs once per task:
human-written rules govern LLM synthesis, provenance gates reject unverifiable
items, and the checklist is frozen before any trajectory is read. During
per-trajectory scoring, a judge labels each item from the execution log
alone; the official outcome is applied only afterwards, at one scripted step.
Aggregation produces the three evaluation metrics: checklist partial credit,
Perfect, and evidence coverage.}
\Description{The GCPC pipeline has two horizontal bands separated by the point
at which the checklist becomes frozen. The upper band synthesizes a checklist
from the task instruction and verifier source under human-written rules, checks
item provenance, and shows a concrete item example. The lower band
judges each item from the frozen checklist and execution log, applies the
official outcome only after item judgment, and aggregates the results into
checklist partial credit, Perfect, and evidence coverage.}
\label{fig:gcpc-pipeline}
\end{figure*}

Fine-grained trajectory evaluation faces a scalability--trustworthiness
dilemma. Expert-written checklists can closely reflect individual tasks, but lack scalability and transferability due to the high authoring cost. Automatically generated checklists scale more easily, but their coverage, granularity, and grounding may be unreliable. To address this issue, GCPC separates the parts
that require human governance from those that benefit from model-scale
instantiation. This separation provides task-specific items without manual per-task checklist authoring, while preventing trajectory-specific adaptation and keeping both checklist synthesis and scoring auditable. Specifically, GCPC organizes evaluation into two stages: \textbf{checklist synthesis} and \textbf{trajectory scoring}. The overall pipeline is illustrated in Figure~\ref{fig:gcpc-pipeline}. 

A task $t$ provides a natural-language instruction $I_t$ and an official
verifier $V_t$ whose executable assertions define what the benchmark accepts.
A public corpus provides trajectories $\tau$ with execution logs and official
PASS/FAIL outcomes. For \textbf{Checklist synthesis}, human experts define a five-category scheme for auditing requirement coverage, an item contract specifying how each item should be written, provenance rules specifying where each item comes from and how it can be verified, and mechanical gates that automatically reject invalid items. An LLM then applies these fixed rules to $(I_t,V_t)$ to generate a task-specific checklist $C_t=\{c_1,\ldots,c_m\}$ (\S\ref{sec:synthesis}). Note that \textbf{checklist synthesis} is performed once per task and generates a frozen checklist with auditable provenance.


\textbf{Trajectory scoring} applies the obtained checklist $C_t$ to each trajectory $\tau$ of the task, judging each item only from evidence visible in the execution log, before incorporating the official outcome at one scripted step (\S\ref{sec:scoring}). The judged items are finally aggregated into the evaluation metrics, i.e., the GCPC score \S\ref{sec:metrics}. Throughout this paper, \emph{GCPC} names both the synthesis-and-scoring
pipeline shown in Figure~\ref{fig:gcpc-pipeline} and its obtained metric in Equation~\ref{eq:cpc}. GCPC is always computed from the execution log alone without the official outcome, while \emph{GCPC with official outcome} denotes the variant that additionally applies the official-outcome step of Section~\ref{sec:scoring}.\\

\subsection{Checklist Synthesis}
\label{sec:synthesis}

The top panel of Figure~\ref{fig:gcpc-pipeline} shows the one-time \textbf{checklist synthesis} stage. Synthesis uses only the task instruction and official verifier, without access to any evaluated trajectory, official outcome, or oracle solution. Excluding the oracle avoids privileging a single reference solution, while freezing the checklist before scoring prevents trajectory-specific adaptation.

\paragraph{Reusable human-governed rules.}
Rather than asking experts to write a checklist for every task,  \textbf{checklist synthesis} of GCPC defines once the rules that a valid item should follow. These reusable rules include an item contract specifying how each item should be written, provenance requirements specifying how it must be grounded in the task, and mechanical validation gates for rejecting invalid items. An LLM then applies the same rules to each task instruction and official verifier to generate a task-specific checklist. Separately, we use a five-category scheme covering comprehension, acquisition, computation, construction, and validation to audit whether the resulting checklist captures the main types of task requirements. This scheme is used only for coverage and failure analysis and does not guide checklist synthesis.

\paragraph{Grounded instantiation and validation.}
A generator model (Claude Sonnet 4.6) applies the fixed rules to the task instruction and official verifier to produce a task-specific checklist. Each checklist consists of self-contained, non-redundant items that apply across valid solution paths, with each item carrying a verbatim quote from the task instruction. Each item is assigned one of three verification routes: \emph{assertion} when a published
verifier test covers it, \emph{generated} when a deterministic check could be
written but is absent and requires a newly generated deterministic check, and \emph{judge} when judgment is unavoidable. Across 87 tasks, the resulting checklists contain 713 items (4--15 per task; median 8), of which
82.3\%, 14.9\%, and 2.8\% follow the \emph{assertion}, \emph{generated}, and \emph{judge} routes, respectively.
Figure~\ref{fig:checklist-examples} shows complete five-item checklists for
two tasks from different domains, illustrating how the same rules produce
task-specific items across all three verification routes.

A deterministic validator then checks each item to ensure that its quoted text appears verbatim in the task instruction, any assertion reference resolves to an actual verifier test, and the item expresses a task requirement rather than merely reproducing test code. All 713 production items pass these gates. Failed generations are explicitly reviewed and regenerated rather than repaired through an unlogged retry loop. Targeted author review is reserved for human-evaluation tasks and items flagged by the enforceability analysis in \S\ref{sec:reliability}.

\begin{figure*}[t]
\centering

\begin{minipage}[t]{0.487\textwidth}
\begin{tcolorbox}[
  enhanced,
  equal height group=checklistcards,
  colback=white,
  colframe=ChkGrayLine,
  boxrule=0.55pt,
  arc=3pt,
  left=5pt,
  right=5pt,
  top=2pt,
  bottom=2pt,
  title={\textbf{Task:Lake Mendota simulation}},
  colbacktitle=ChkBlueBg,
  coltitle=ChkBlueText,
  fonttitle=\sffamily\normalsize,
  boxed title style={
    colframe=ChkBlueLine,
    boxrule=0pt,
    arc=2pt
  },
  borderline north={1.2pt}{0pt}{ChkBlueLine}
]

\centering
{\sffamily\scriptsize
\texttt{glm-lake-mendota}
\quad $\cdot$ \quad
5 items
\quad $\cdot$ \quad
5/5 mechanically valid
}
\par\vspace{2pt}

\renewcommand{\arraystretch}{1.02}
\begin{tabularx}{\linewidth}{
  @{}
  >{\raggedright\arraybackslash}p{0.12\linewidth}
  >{\raggedright\arraybackslash}X
  >{\raggedleft\arraybackslash}p{0.23\linewidth}
  @{}
}
\textbf{\scriptsize ID}
  & \textbf{\scriptsize Generated item}
  & \textbf{\scriptsize Route} \\
\arrayrulecolor{ChkRule}\hline

{\scriptsize GLM-01}
  & {\scriptsize Run GLM successfully to simulate Lake Mendota's
    vertical water temperature.}
  & \assertionroute{assertion} \\
\hline

{\scriptsize GLM-02}
  & {\scriptsize Achieve temperature RMSE below $2.0^{\circ}$C against the
    observations.}
  & \assertionroute{assertion} \\
\hline

{\scriptsize GLM-03}
  & {\scriptsize Produce the simulation output as
    \texttt{/root/output/output.nc}.}
  & \assertionroute{assertion} \\
\hline

{\scriptsize GLM-04}
  & {\scriptsize Cover the complete requested period,
    2009-01-01 through 2015-12-30.}
  & \generatedroute{generated} \\
\hline

{\scriptsize GLM-05}
  & {\scriptsize Save the calibrated parameters to
    \texttt{/root/glm3.nml} so that GLM can be rerun standalone.}
  & \assertionroute{assertion} \\

\end{tabularx}
\end{tcolorbox}
\end{minipage}
\hfill
\begin{minipage}[t]{0.487\textwidth}
\begin{tcolorbox}[
  enhanced,
  equal height group=checklistcards,
  colback=white,
  colframe=ChkGrayLine,
  boxrule=0.55pt,
  arc=3pt,
  left=5pt,
  right=5pt,
  top=2pt,
  bottom=2pt,
  title={\textbf{Task:Diagnose and repair a failed build}},
  colbacktitle=ChkPinkBg,
  coltitle=ChkPinkText,
  fonttitle=\sffamily\normalsize,
  boxed title style={
    colframe=ChkPinkLine,
    boxrule=0pt,
    arc=2pt
  },
  borderline north={1.2pt}{0pt}{ChkPinkLine}
]

\centering
{\sffamily\scriptsize
\texttt{fix-build-google-auto}
\quad $\cdot$ \quad
5 items
\quad $\cdot$ \quad
5/5 mechanically valid
}
\par\vspace{2pt}

\renewcommand{\arraystretch}{1.02}
\begin{tabularx}{\linewidth}{
  @{}
  >{\raggedright\arraybackslash}p{0.12\linewidth}
  >{\raggedright\arraybackslash}X
  >{\raggedleft\arraybackslash}p{0.23\linewidth}
  @{}
}
\textbf{\scriptsize ID}
  & \textbf{\scriptsize Generated item}
  & \textbf{\scriptsize Route} \\
\arrayrulecolor{ChkRule}\hline

{\scriptsize S-01}
  & {\scriptsize Save a non-empty analysis of the build failure to
    \texttt{failed\_reasons.txt}.}
  & \assertionroute{assertion} \\
\hline

{\scriptsize S-02}
  & {\scriptsize Name the concrete errors and classify each as originating
    in application code or build configuration.}
  & \judgeroute{judge} \\
\hline

{\scriptsize S-03}
  & {\scriptsize Produce one or more
    \texttt{patch\_\{i\}.diff} files in standard unified-diff format.}
  & \assertionroute{assertion} \\
\hline

{\scriptsize S-04}
  & {\scriptsize Ensure that the proposed patches address the errors
    identified in the failure analysis.}
  & \judgeroute{judge} \\
\hline

{\scriptsize S-05}
  & {\scriptsize Apply the patches so that the official build script
    subsequently exits with code 0.}
  & \assertionroute{assertion} \\

\end{tabularx}
\end{tcolorbox}
\end{minipage}

\caption{\textbf{Examples of frozen task-specific checklists.}
All items are shown (wording shortened for layout). Routes mark requirements
covered by an existing verifier test (\textsc{assertion}), admitting a new
deterministic check (\textsc{generated}), or requiring model judgment
(\textsc{judge}). Each checklist is synthesized once per task, validated, and
reused across trajectories, without seeing any trajectory or outcome.}
\label{fig:checklist-examples}
\end{figure*}
\subsection{Trajectory Scoring}
\label{sec:scoring}

The lower subfigure of Figure~\ref{fig:gcpc-pipeline} shows how the checklist is applied to an execution trajectory, while keeping trajectory-visible evidence separate from information available only to the official verifier.

\paragraph{Evidence-grounded item judgment.}
A judge model (GPT-5.4-mini, from a different model family than the generator)
receives the checklist and one execution log, and independently labels
each item \textsc{Yes}, \textsc{No}, or \abstain{}. Only displayed tool
outputs and file contents count as evidence; agent claims and unobserved state
do not. \abstain{} means that the log does not contain enough evidence to decide whether the item is satisfied or not, rather than that the requirement has failed. This item-level judgment keeps scoring tied to observable evidence and makes missing evidence explicit.

\paragraph{Applying the official outcome.}
Only after every item has been judged does a deterministic script incorporate the
official outcome. On an official PASS, \emph{PASS propagation} overwrites
assertion-route items as \textsc{Yes}, because the verifier has executed
the referenced tests against the saved artifact. On an official FAIL, the
item judgments remain unchanged because the endpoint does not reveal
which individual requirements failed. The script then appends one virtual
\emph{official-outcome endpoint item}, labeled \textsc{Yes} for PASS and
\textsc{No} for FAIL. This endpoint contributes to \emph{GCPC with official outcome}, ensuring that the partial-credit score remains consistent with final task success, but is excluded from Perfect. For example, suppose all nine checklist items are judged \textsc{Yes} from the execution log, but the official verifier returns FAIL. Plain GCPC, which uses only log evidence, assigns a score of $9/9=1.0$.  \emph{GCPC with official outcome} additionally includes a \textsc{No} endpoint for the failed task, giving $9/(9+1)=0.900$. Thus, a failed trajectory may still receive high partial credit when the log shows substantial progress, but it cannot receive a perfect score once the official outcome is incorporated. Plain \emph{GCPC} uses only trajectory evidence and does not incorporate the official PASS/FAIL outcome. In particular, it applies neither PASS propagation nor the official-outcome endpoint item. We use this log-only score for the equal-information comparisons in \S\ref{sec:experiments}, while later ablations separately measure the effects of these two official-outcome mechanisms.


\subsection{Evaluation Metrics}
\label{sec:metrics}

For a trajectory $\tau$, the judgment stage assigns every item
$c_i\in C_t$ one of \textsc{Yes}, \textsc{No}, or \abstain{}. Let $D_\tau$
denote the items decided after the official-outcome step
(\S\ref{sec:scoring}). The GCPC pipeline reports three complementary metrics.

\paragraph{Checklist Partial Credit.}
The primary checklist partial credit score measures the fraction of criteria that are satisfied:
\begin{equation}
\label{eq:cpc}
\cpc{}(\tau)=
\frac{|\{c\in D_\tau:\operatorname{label}(c)=\textsc{Yes}\}|}
{|D_\tau|}.
\end{equation}
Criteria labeled \abstain{} are excluded from the denominator, so insufficient
evidence is not treated as failure. The amount of checklist evidence supporting
the score is reported separately through evidence coverage.
\paragraph{Zero-violation companion.}
A high partial-credit score can still hide one fatal violation among many satisfied items. We therefore report a complementary binary metric 
\begin{equation}
\mathrm{Perfect}(\tau)=
\mathbf{1}\!\left[\nexists c\in C_t:
\operatorname{label}(c)=\textsc{No}\right].
\end{equation}
The Perfect score equals one if no checklist item is judged
\textsc{No}, and zero otherwise. Items labeled \abstain{} do not count as
violations. The metric is computed only over the task checklist and excludes
the virtual official-outcome endpoint introduced in \S\ref{sec:scoring}.
Including this endpoint would make the results largely duplicate the
official PASS/FAIL outcome: On the deduplicated $n{=}4{,}455$ population,
Perfect agrees with the official outcome on 97.6\% of trajectories when the endpoint is included, but on only 66.4\% when it is excluded.

\paragraph{Evidence coverage.}
We report $|D_\tau|/|C_t|$ as the fraction of checklist criteria for which the execution log provides sufficient evidence to determine whether the corresponding checklist is satisfied. If all criteria receive \abstain{}, the trajectory is marked as zero-evidence and handled separately in the experimental protocol (\S\ref{sec:experiments}).

\section{Experiments}
\label{sec:experiments}

We organize our experiments to validate the proposed evaluation pipeline around five research questions (RQs). \textbf{RQ1} tests whether checklist
partial credit extracts more signal from a trajectory than existing scoring
methods under equal information. \textbf{RQ2} evaluates whether that signal
agrees with independent human judgments of execution progress. \textbf{RQ3}
asks what the score reveals about skills when the official outcome does not
change. \textbf{RQ4} asks which constraints are needed to make scalable
checklist evaluation trustworthy, isolating task grounding, the
official-outcome step, and abstention. Finally, \textbf{RQ5} tests whether the
instrument is repeatable, enforceable, robust to judge substitution, and
applicable beyond the tasks used to develop it.

\subsection{Experimental Setup}
\label{sec:setup}

\paragraph{Benchmark.}
We study SkillsBench v1.1, which contains 87 tool-using tasks from eight domains
and 20{,}818 labeled public trajectories produced by 234
harness--model--condition configurations. Each trajectory consists of an
execution log and an official PASS/FAIL outcome. Our
data collection covers all tasks with 713 items (4--15 per
task; median 8), all of which pass the mechanical provenance checks of
\S\ref{sec:synthesis}. The evaluation population contains 4{,}455 unique
trajectories with binary official outcomes; self-generated-skill
configurations and non-binary rewards are excluded wherever a matched binary
comparison is required. Answering the above different RQs requires different stored judgments or matching
constraints. Each population is the largest or most appropriate subset for a particular analysis, depending on whether the analysis requires the same trajectories to be scored by both methods, repeated criterion judgments, item-level judgments for the official-outcome-step ablation, matched with/without-skill pairs, or human annotations.




\paragraph{Compared scoring methods and settings.}
We compare five log-only scoring methods: \emph{GCPC} (plain GCPC without the official-outcome step in this comparison), a \emph{Generic checklist}, \emph{TICK} as published \citep{cook2024tick}, an \emph{adapted TICK}, and a \emph{Holistic judge}. All five methods receive the same trajectory information and do not see the official PASS/FAIL outcome. All methods that require model-based judgment use the same judge model, so the comparison varies the scoring procedure while controlling for judge identity.

The \emph{Generic checklist} uses the same five fixed questions for all tasks, without adaptation to the task instruction or verifier. This baseline tests whether using multiple criteria alone improves trajectory evaluation. \emph{TICK} follows the original checklist generation and scoring procedure of \citet{cook2024tick}. Since TICK was designed for evaluating responses, we also test an \emph{adapted TICK} that replaces the word \emph{response} with \emph{agent execution trajectory} while leaving the rest of the procedure unchanged. The \emph{Holistic judge} evaluates the entire execution trajectory with a single score from 0 to 10, which is rescaled to $[0,1]$. It serves as a baseline for comparing criterion-level scoring with overall trajectory-level scoring.

Metrics that use the official outcome, including \emph{GCPC with official outcome}, \emph{Perfect}, and \emph{Pass Rate}, are analyzed separately because they incorporate the same PASS/FAIL outcome against which discrimination is measured.

\paragraph{Metrics and uncertainty.}
For agreement with the official outcome, we report area under the ROC curve
(AUC), with ties counted as one half, and PASS--FAIL mean separation where a
mechanism analysis requires an additive scale. Confidence intervals use
2{,}000 task-clustered bootstrap resamples with fixed seeds. These measures assess agreement with the official binary outcome, but do not establish agreement with human judgments. Human validity is evaluated separately using pairwise preference agreement and
Spearman correlation with mean progress ratings. For outcome-free analyses, trajectories for which all criteria receive \abstain{} are excluded, while individual abstained criteria are excluded from the GCPC denominator.

\subsection{RQ1: Does GCPC improve trajectory discrimination?}
\label{sec:main}

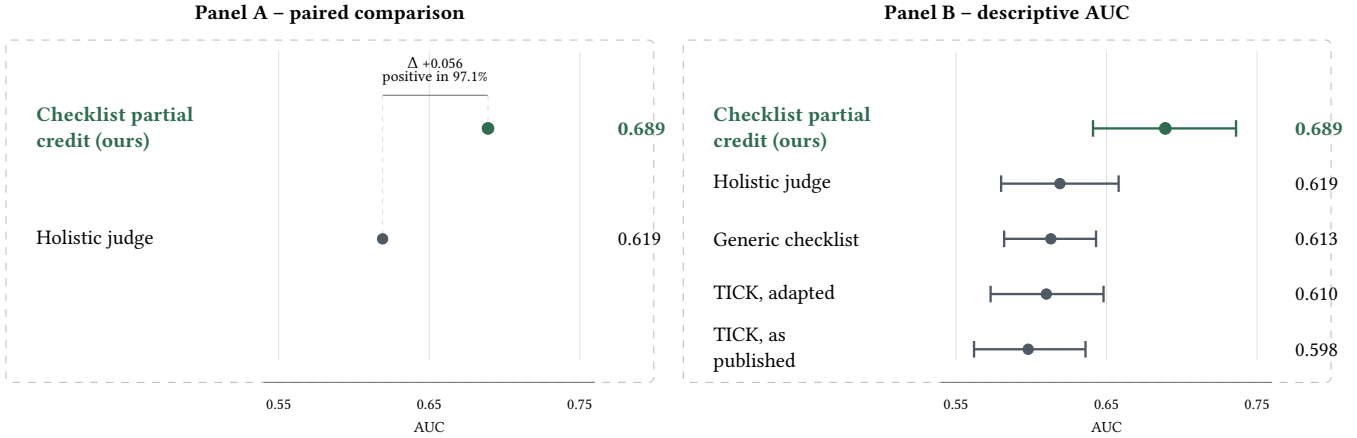
\begin{figure*}[t]
\centering
\resizebox{\textwidth}{!}{%
\begin{tikzpicture}[
  yscale=0.75,
  every node/.style={font=\small},
  gridline/.style={draw=ChkRule, line width=0.3pt},
  axisline/.style={draw=black!55, line width=0.4pt},
  whisker/.style={draw=ACEGrayText, line width=0.9pt},
  whiskerours/.style={draw=acepos, line width=0.9pt},
]

\begin{scope}[xshift=0cm]
  \draw[draw=ChkGrayLine, dashed, line width=0.4pt, rounded corners=2pt]
    (-0.3,0.6) rectangle (8.5,-5.6);
  \node[font=\small\bfseries, anchor=south] at (4.1,0.75) {Panel A -- paired comparison};

  \def\Aorigin{3.2}
  \foreach \v in {0.205,2.250,4.295}{
    \draw[gridline] (\Aorigin+\v,0.4) -- (\Aorigin+\v,-5.2);
  }

  \draw[axisline] (\Aorigin+1.616,-0.4) -- (\Aorigin+3.048,-0.4);
  \draw[ChkRule, dash pattern=on 2pt off 2pt] (\Aorigin+1.616,-0.4) -- (\Aorigin+1.616,-3.0);
  \draw[ChkRule, dash pattern=on 2pt off 2pt] (\Aorigin+3.048,-0.4) -- (\Aorigin+3.048,-1.0);
  \node[align=center, font=\scriptsize] at (\Aorigin+2.33,0.05)
    {$\Delta$ +0.056\\[-2pt]positive in 97.1\%};

  \node[draw=acepos, fill=acepos, circle, minimum size=4.5pt, inner sep=0pt] at (\Aorigin+3.048,-1.0) {};
  \node[anchor=west, align=left, font=\small\bfseries, text=acepos] at (0,-1.0) {Checklist partial\\credit (ours)};
  \node[anchor=west, font=\small\bfseries, text=acepos] at (\Aorigin+4.7,-1.0) {0.689};

  \node[draw=ACEGrayText, fill=ACEGrayText, circle, minimum size=4pt, inner sep=0pt] at (\Aorigin+1.616,-3.0) {};
  \node[anchor=west, font=\small] at (0,-3.0) {Holistic judge};
  \node[anchor=west, font=\small] at (\Aorigin+4.7,-3.0) {0.619};

  \draw[axisline] (\Aorigin,-5.6) -- (\Aorigin+4.5,-5.6);
  \node[font=\scriptsize] at (\Aorigin+0.205,-6.0) {0.55};
  \node[font=\scriptsize] at (\Aorigin+2.250,-6.0) {0.65};
  \node[font=\scriptsize] at (\Aorigin+4.295,-6.0) {0.75};
  \node[font=\scriptsize] at (\Aorigin+2.25,-6.4) {AUC};
\end{scope}

\begin{scope}[xshift=9.2cm]
  \draw[draw=ChkGrayLine, dashed, line width=0.4pt, rounded corners=2pt]
    (-0.3,0.6) rectangle (8.5,-5.6);
  \node[font=\small\bfseries, anchor=south] at (4.1,0.75) {Panel B -- descriptive AUC};

  \def\Borigin{3.2}
  \foreach \v in {0.205,2.250,4.295}{
    \draw[gridline] (\Borigin+\v,0.4) -- (\Borigin+\v,-5.2);
  }

  \draw[whiskerours] (\Borigin+2.066,-1.0) -- (\Borigin+4.009,-1.0);
  \draw[whiskerours] (\Borigin+2.066,-1.15) -- (\Borigin+2.066,-0.85);
  \draw[whiskerours] (\Borigin+4.009,-1.15) -- (\Borigin+4.009,-0.85);
  \node[draw=acepos, fill=acepos, circle, minimum size=4.5pt, inner sep=0pt] at (\Borigin+3.048,-1.0) {};
  \node[anchor=west, align=left, font=\small\bfseries, text=acepos] at (0,-1.0) {Checklist partial\\credit (ours)};
  \node[anchor=west, font=\small\bfseries, text=acepos] at (\Borigin+4.7,-1.0) {0.689};

  \draw[whisker] (\Borigin+0.818,-2.0) -- (\Borigin+2.414,-2.0);
  \draw[whisker] (\Borigin+0.818,-2.15) -- (\Borigin+0.818,-1.85);
  \draw[whisker] (\Borigin+2.414,-2.15) -- (\Borigin+2.414,-1.85);
  \node[draw=ACEGrayText, fill=ACEGrayText, circle, minimum size=4pt, inner sep=0pt] at (\Borigin+1.616,-2.0) {};
  \node[anchor=west, font=\small] at (0,-2.0) {Holistic judge};
  \node[anchor=west, font=\small] at (\Borigin+4.7,-2.0) {0.619};

  \draw[whisker] (\Borigin+0.859,-3.0) -- (\Borigin+2.107,-3.0);
  \draw[whisker] (\Borigin+0.859,-3.15) -- (\Borigin+0.859,-2.85);
  \draw[whisker] (\Borigin+2.107,-3.15) -- (\Borigin+2.107,-2.85);
  \node[draw=ACEGrayText, fill=ACEGrayText, circle, minimum size=4pt, inner sep=0pt] at (\Borigin+1.493,-3.0) {};
  \node[anchor=west, font=\small] at (0,-3.0) {Generic checklist};
  \node[anchor=west, font=\small] at (\Borigin+4.7,-3.0) {0.613};

  \draw[whisker] (\Borigin+0.675,-4.0) -- (\Borigin+2.209,-4.0);
  \draw[whisker] (\Borigin+0.675,-4.15) -- (\Borigin+0.675,-3.85);
  \draw[whisker] (\Borigin+2.209,-4.15) -- (\Borigin+2.209,-3.85);
  \node[draw=ACEGrayText, fill=ACEGrayText, circle, minimum size=4pt, inner sep=0pt] at (\Borigin+1.432,-4.0) {};
  \node[anchor=west, font=\small] at (0,-4.0) {TICK, adapted};
  \node[anchor=west, font=\small] at (\Borigin+4.7,-4.0) {0.610};

  \draw[whisker] (\Borigin+0.450,-5.0) -- (\Borigin+1.964,-5.0);
  \draw[whisker] (\Borigin+0.450,-5.15) -- (\Borigin+0.450,-4.85);
  \draw[whisker] (\Borigin+1.964,-5.15) -- (\Borigin+1.964,-4.85);
  \node[draw=ACEGrayText, fill=ACEGrayText, circle, minimum size=4pt, inner sep=0pt] at (\Borigin+1.186,-5.0) {};
  \node[anchor=west, align=left, font=\small] at (0,-5.0) {TICK, as\\published};
  \node[anchor=west, font=\small] at (\Borigin+4.7,-5.0) {0.598};

  \draw[axisline] (\Borigin,-5.6) -- (\Borigin+4.5,-5.6);
  \node[font=\scriptsize] at (\Borigin+0.205,-6.0) {0.55};
  \node[font=\scriptsize] at (\Borigin+2.250,-6.0) {0.65};
  \node[font=\scriptsize] at (\Borigin+4.295,-6.0) {0.75};
  \node[font=\scriptsize] at (\Borigin+2.25,-6.4) {AUC};
\end{scope}

\end{tikzpicture}%
}
\caption{Checklist partial credit vs. baseline methods ($n{=}4{,}455$). Panel
A: paired comparison with the holistic judge on identical trajectories
(paired $\Delta$AUC $+0.056$, 95\% CI $[-0.002,0.118]$, positive in 97.1\% of
resamples). Panel B: descriptive AUC (95\% CI) on each method's own available
sample.}
\label{tab:main}
\end{figure*}

To isolate the effect of the scoring method from differences in population
coverage, our primary comparison uses the trajectories scored by both
GCPC and the holistic judge, and the corresponding results are presented in Table \ref{tab:main}. On these identical trajectories, GCPC achieves
an AUC of $0.689$, compared with $0.619$ for the holistic judge. The corresponding paired
$\Delta$AUC is $+0.056$ with a task-clustered 95\% confidence interval of
$[-0.002,0.118]$; 97.1\% of the 2{,}000 bootstrap draws are positive. Thus, item-
level scoring discriminates official outcomes more reliably than asking the
same judge model for one trajectory-level score.

The remaining arms probe rival explanations descriptively. The fixed Generic
checklist reaches AUC $0.613$, showing that merely replacing one holistic
question with several fixed questions does not recover GCPC's signal. TICK
reaches $0.598$ as published and $0.610$ after the minimal trajectory-domain
adaptation; the tested adaptation narrows but does not close the gap. Because
these rows have different coverage and differ in more than one design choice,
they identify which broad explanations are insufficient rather than assigning
the entire gain to a single component. GCPC is also stable across
skill condition (AUC $0.681$ with skills and $0.688$ without), indicating that
the result is not carried by only one condition.

Under equal information and on identical trajectories, GCPC provides more
outcome-discriminative signal than holistic judging. This is a mechanism
check against the benchmark label, not yet evidence that GCPC matches how
people perceive partial progress; RQ2 supplies that independent target.

\subsection{RQ2: Does checklist partial credit agree with human judgments?}
\label{sec:humaneval}

We evaluate GCPC against human judgments at three levels by answering three questions, Q1--Q3. Q1 asks humans to rate the overall progress
of each trajectory and tests whether higher GCPC scores correspond to greater
perceived progress. Q2 asks humans to compare trajectories from the same task
and tests whether GCPC ranks better and worse executions in the same way as
humans. Q3 asks humans to judge individual checklist criteria and tests whether
the criterion-level judgments produced by GCPC agree with human judgments.
Q1 and Q2 are completed before annotators see the checklist, providing
independent validation of the GCPC score. Q3 is conducted after the checklist
is revealed and directly evaluates agreement on individual criteria.

We collected annotations for a prespecified subset of 96 trajectories from
12 tasks spanning all eight SkillsBench task domains. Five annotators completed
25 task assignments, with every task assigned to at least two annotators. For
each assignment, an annotator first evaluated all eight same-task trajectories
under Q1/Q1b and Q2. Q3 was then conducted on five selected
trajectories, each evaluated by two annotators, providing 
criterion-level audit. Annotators were not given official outcomes, automated scores, or
configuration metadata, although some logs revealed the harness used.
\begin{table}[t]
\centering
\small
\caption{Agreement with human judgments (pairwise agreement with Q2
preferences; $\rho$ with mean Q1 ratings). Comparable-pair counts vary by
method due to ties and missing scores.}
\label{tab:human}
\begin{tabular}{@{}lll@{}}
\toprule
Method & Pairwise agr. & $\rho$ vs. Q1 [95\% CI] \\
\midrule
GCPC (ours) & \textbf{0.730} & 0.382 [$0.114,0.628$] \\
GCPC with official outcome (ours) & 0.656 & 0.384 [$0.111,0.642$] \\
Holistic judge & 0.613 & 0.218 [$0.067,0.390$] \\
TICK & 0.691 & 0.162 (n.s.) \\
Generic checklist & 0.569 & 0.132 (n.s.) \\
\bottomrule
\end{tabular}
\end{table}

Among the continuous outcome-free methods, GCPC achieves the highest agreement with
human pairwise preferences, reaching $0.730$ compared with $0.613$ for
the holistic judge, $0.691$ for TICK, and $0.569$ for the Generic checklist. GCPC also shows a stronger correlation with human progress ratings than the
holistic judge ($\rho=0.382$ versus $0.218$). 

Human annotations further reveal two challenges in trajectory evaluation.
First, holistic progress ratings show only moderate inter-annotator agreement
($\alpha=0.466$), whereas criterion-level judgments reach $\alpha=1.00$ on
the small doubly rated subset of 12 criteria. Although the latter result is
based on a limited sample, it suggests that individual requirements may be
easier to judge consistently than overall execution progress. Second, when
annotators see only the execution log, they assign similar progress scores to
officially passing and failing trajectories ($8.86/10$ versus $7.88/10$ on
average). This indicates that the log does not always reveal the final
artifact state available to the official verifier, motivating the separate
official-outcome step. Because this step directly uses the official outcome,
\emph{GCPC with official outcome} is not treated as independent evidence of human
validity.


\subsection{RQ3: What skill effects does binary evaluation hide?}
\label{sec:skilleffect}

To measure skill-associated changes, we construct matched pairs that share the
same task, harness, and model-and-source configuration, and differ only in
whether a skill is provided. After excluding self-generated-skill
configurations and non-binary outcomes, we obtain 1{,}946 matched pairs. For
each pair, we compute
$\Delta\cpc{}=\cpc{}_{\mathrm{with}}-\cpc{}_{\mathrm{without}}$, where a
positive value indicates higher checklist partial credit with the skill. We then group the pairs by how the official PASS/FAIL outcome changes after
adding the skill: FAIL$\rightarrow$FAIL, FAIL$\rightarrow$PASS,
PASS$\rightarrow$FAIL, or PASS$\rightarrow$PASS. The matching rule, the
$0.037$ test--retest noise floor, the $0.10$ threshold for a material change,
and the statistical tests were fixed before computing the results.

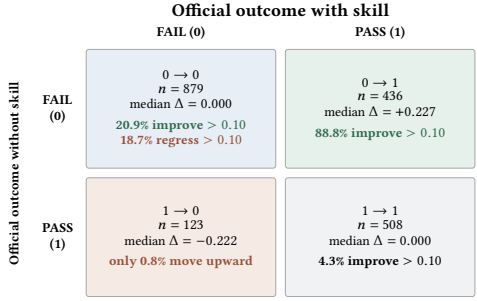
\begin{figure}[t]
\centering

\scalebox{0.85}{\begin{tikzpicture}[
    data/.style={
        font=\scriptsize,
        align=center
    },
    cell/.style={
        data,
        draw=black!35,
        rounded corners=1.2pt,
        minimum width=2.95cm,
        minimum height=1.85cm,
        inner sep=2.5pt
    }
]


\node[font=\small\bfseries] at (3.05,2.55)
    {Official outcome with skill};

\node[font=\scriptsize\bfseries] at (1.50,2.18)
    {FAIL (0)};

\node[font=\scriptsize\bfseries] at (4.60,2.18)
    {PASS (1)};


\node[font=\scriptsize\bfseries, rotate=90] at (-1.10,0)
    {Official outcome without skill};

\node[font=\scriptsize\bfseries, align=center] at (-0.42,1.00)
    {FAIL\\(0)};

\node[font=\scriptsize\bfseries, align=center] at (-0.42,-1.00)
    {PASS\\(1)};


\node[cell, fill=aceblue] at (1.50,1.00) {
    \textbf{$0\rightarrow0$}\\[-1pt]
    $n=879$\\
    median $\Delta=0.000$\\[2pt]
    \textcolor{acepos}{\textbf{20.9\% improve $>0.10$}}\\
    \textcolor{aceneg}{\textbf{18.7\% regress $>0.10$}}
};


\node[cell, fill=acegreen] at (4.60,1.00) {
    \textbf{$0\rightarrow1$}\\[-1pt]
    $n=436$\\
    median $\Delta=+0.227$\\[2pt]
    \textcolor{acepos}{\textbf{88.8\% improve $>0.10$}}
};


\node[cell, fill=aceorange] at (1.50,-1.00) {
    \textbf{$1\rightarrow0$}\\[-1pt]
    $n=123$\\
    median $\Delta=-0.222$\\[2pt]
    \textcolor{aceneg}{\textbf{only 0.8\% move upward}}
};


\node[cell, fill=acegray] at (4.60,-1.00) {
    \textbf{$1\rightarrow1$}\\[-1pt]
    $n=508$\\
    median $\Delta=0.000$\\[2pt]
    \textbf{4.3\% improve $>0.10$}
};

\end{tikzpicture}}

\caption{Matched skill effects by official-outcome transition
(without $\rightarrow$ with skill).} 

\label{fig:skill-transitions}
\end{figure}

Figure~\ref{fig:skill-transitions} compares changes in the official PASS/FAIL
outcome with changes in checklist partial credit. When adding a skill changes
the official outcome, GCPC moves in the expected direction. For
FAIL$\rightarrow$PASS pairs, the median GCPC increases by $0.227$, whereas
for PASS$\rightarrow$FAIL pairs, it decreases by $0.222$.

The largest group consists of 879 FAIL$\rightarrow$FAIL pairs, for which
binary evaluation reports no change. Notably, 20.9\% of pairs improve by more than $0.10$, while 18.7\% regress by more than $0.10$. Thus, even when the official outcome remains unchanged, GCPC reveals substantial skill-associated improvements and
regressions and identifies which checklist criteria changed.


\subsection{RQ4: Which design choices make GCPC trustworthy?}
\label{sec:ablation}

We next examine which design choices are important for reliable GCPC evaluation. Specifically, we test whether the checklist should be task-specific, how the official PASS/FAIL outcome should be incorporated, how insufficient evidence should be handled, and whether criterion judgments actually respond to the evidence they are intended to assess. We also evaluate the five-category requirement scheme used to audit checklist coverage. Table~\ref{tab:design-ablation} summarizes these analyses.

\begin{table*}[t]
\centering
\footnotesize
\setlength{\tabcolsep}{4.2pt}
\renewcommand{\arraystretch}{1.10}

\caption{Ablations and falsification checks. Panel A reports descriptive
AUCs on each arm's own available sample within the n=4{,}455 deduplicated
population; Panel B compares the same 3{,}811 board trajectories with recovered item-level judgments. Panel C reports
separate quality-control checks.}
\label{tab:design-ablation}

\begin{tabular}{
    @{}
    >{\raggedright\arraybackslash}p{0.19\textwidth}
    >{\raggedright\arraybackslash}p{0.35\textwidth}
    >{\raggedright\arraybackslash}p{0.13\textwidth}
    >{\raggedleft\arraybackslash}p{0.22\textwidth}
    @{}
}
\toprule
\textbf{Variant or test}
    & \textbf{Design change}
    & \textbf{Sample}
    & \textbf{Result} \\
\midrule

\rowcolor{ACEBlueBand}
\multicolumn{4}{@{}l@{}}{
    \hspace{4pt}
    \textcolor{ACEBlueText}{
        \textbf{A \quad Checklist construction}
        \hfill
        \textnormal{\emph{descriptive AUC}}
    }
} \\
\addlinespace[1pt]

\textbf{Task-specific GCPC (ours)}
    & Grounded checklist generated separately for each task
    &
    & \cellcolor{ACEBlueCell}
      \textcolor{ACEBlueText}{\textbf{0.689}} \\

Generic checklist
    & Replace task-specific items with five fixed questions
    &
    & 0.613 \\

TICK
    & Replace checklist synthesis with an external pipeline
    &
    & 0.598 \\

TICK, adapted
    & Add trajectory-domain wording to the TICK prompt
    &
    & 0.610 \\

\addlinespace[3pt]

\rowcolor{ACEPinkBand}
\multicolumn{4}{@{}l@{}}{
    \hspace{4pt}
    \textcolor{ACEPinkText}{
        \textbf{B \quad The official-outcome step
        }
        \hfill
        \textnormal{\emph{PASS--FAIL separation}}
    }
} \\
\addlinespace[1pt]

GCPC (no official outcome)
    & Endpoint off; PASS propagation off
    &
    & 0.121 \\

Endpoint only
    & Endpoint on; PASS propagation off
    &
    & 0.315 \quad
      \textcolor{ACEBlueText}{$(+0.194)$} \\

Propagation only
    & Endpoint off; PASS propagation on
    &
    & 0.155 \quad
      \textcolor{ACEBlueText}{$(+0.034)$} \\

\textbf{Both operations}
    & Endpoint on; PASS propagation on
    &
    & \cellcolor{ACEBlueCell}
      \textcolor{ACEBlueText}{
          \textbf{0.338} \quad $\mathbf{(+0.217)}$
      } \\

\addlinespace[3pt]

\rowcolor{ACEGrayBand}
\multicolumn{4}{@{}l@{}}{
    \hspace{4pt}
    \textcolor{ACEGrayText}{
        \textbf{C \quad Stress and falsification checks}
        \hfill
        \textnormal{\emph{separate QC checks}}
    }
} \\
\addlinespace[1pt]

Primary abstention rule
    & Exclude \abstain{} from the GCPC denominator
    & 3,811 board traj.
    & Separation 0.338 \\

Forced-verdict sensitivity
    & Recode every \abstain{} as \textsc{No}
    & 3,811 board traj.
    & \cellcolor{ACEPinkCell}
      \textcolor{ACEPinkText}{\textbf{Separation 0.445}} \\

Evidence replacement
    & Replace every occurrence of the decisive evidence
    & 218 item pairs
    & \cellcolor{ACEBlueCell}
      \textcolor{ACEBlueText}{
          \textbf{73.5\%} strict;
          \textbf{78.1\%} incl.\ \abstain{}
      } \\

Idle-agent floor
    & Score trajectories with no substantive task execution
    & 4 traj.
    & 4/4: GCPC $=0$ \\

Schema falsification
    & Real / reduced / irrelevant / name-permuted schema
    & Requirement sets
    & 99.4 / 79.6 / 0.0 / 98.7\% \\

\bottomrule
\end{tabular}
\end{table*}
\paragraph{Checklist construction.}
Panel A tests whether task-specific checklist construction is necessary. GCPC achieves the best performance with an AUC of $0.689$. Replacing its task-specific checklist with the same five fixed questions for every task reduces AUC to $0.613$. TICK reaches $0.598$ using its published procedure and $0.610$ after adapting its prompt to agent trajectories. Because score availability differs across these methods, Panel A provides descriptive comparisons, and the direct paired comparisons on identical trajectories are reported separately in Table~\ref{tab:main}. 
\paragraph{Official-outcome step.}
Panel B isolates the two operations in the official-outcome step, recomputed
at board scale (3{,}811 trajectories with recovered item-level judgments;
110-trajectory directed-set numbers, $0.088/0.222/0.121/0.249$, replicate the
same ordering and are retained as the original replication point). The official
endpoint supplies the larger increment in PASS--FAIL separation
($+0.194$), while PASS propagation alone contributes $+0.034$; combining
them yields $0.338$. These values explain how the official outcome affects
the outcome-inclusive score and are not independent validity evidence. Accordingly,
the primary equal-information comparison in Table~\ref{tab:main} uses plain GCPC.

\paragraph{Abstention and evidence sensitivity.}
Panel C examines whether criterion judgments are appropriately tied to the
available evidence. We first test how missing evidence should be handled.
If every \abstain{} is recoded as \textsc{No}, PASS--FAIL separation increases
from $0.338$ to $0.445$. However, this improvement comes from treating
insufficient evidence as evidence of failure. We therefore retain the original
rule, under which \abstain{} indicates that the requirement cannot be determined
from the log and is excluded from the GCPC denominator.

We next test whether a criterion judgment changes when its supporting evidence
changes. For criteria initially judged \textsc{Yes}, we replace the decisive
evidence with a plausible but incorrect value and score the criterion again.
After replacement, $73.5\%$ of judgments change to \textsc{No}, and $78.1\%$
change to either \textsc{No} or \abstain{}. This shows that most criterion
judgments are sensitive to the evidence in the execution log rather than
remaining unchanged when that evidence becomes incorrect.

\paragraph{Audit-layer falsification.}
The five-category requirement scheme is used only to audit checklist coverage. We compare the original scheme with three controls: an incomplete version with some category definitions removed, an irrelevant version with unrelated definitions, and a name-permuted version that changes category names while preserving their definitions. Coverage decreases from 99.4\% with the original scheme to 79.6\% with the incomplete scheme and 0.0\% with the irrelevant scheme, while the name-permuted version retains 98.7\% coverage. These results show that the audit depends on the completeness and semantic content of the category definitions rather than their names.


\subsection{RQ5: Is GCPC reliable and transferable?}
We finally examine whether GCPC produces stable scores and whether the same
pipeline can be applied beyond skill-conditioned evaluation.

\label{sec:reliability}

\paragraph{Reliability.}
The scoring pipeline runs three times over the 116-trajectory directed set, yielding Krippendorff's $\alpha=0.905$ (95\% CI $[0.811,0.963]$). To check whether this
reliability result generalizes beyond the directed set's composition, we
repeated the same three-pass procedure on an independently sampled set of 60
tasks drawn from a population with much higher skill usage, obtaining
$\alpha=0.937$ [$0.875,0.981$]. Replacing the judge
with Claude Haiku 4.5 gives Spearman $\rho=0.615$ for GCPC and
$\rho=0.916$ for outcome-inclusive scores, although the latter partly reflects the
official signal shared by both judges. Detailed stability strata and
judge-specific abstention behavior are reported in the supplementary
material.

\paragraph{Zero-adaptation transfer.}
Terminal-Bench 2.0 \citep{merrill2026terminalbench} and SWE-bench Verified
\citep{jimenez2024swebench} contain no skill condition. We therefore use them
to test whether GCPC can evaluate general agent trajectories without changing
the generation prompt, item contract, judge prompts, or scoring rules. We
apply the GCPC to one successful and one unsuccessful trajectory for each of 86 Terminal-Bench tasks and 100 SWE-bench Verified
instances, and present the results in Table \ref{tab:crossbench-main}.

\begin{table}[t]
\centering
\small
\caption{Zero-adaptation checklist synthesis across benchmarks.}
\label{tab:crossbench-main}
\begin{tabular}{@{}lrrr@{}}
\toprule
Benchmark & Tasks & Checklist Items & Validity \\
\midrule
SkillsBench v1.1      & 87  & 713 & 100\% \\
Terminal-Bench 2.0    & 86  & 585 & 98.8\% \\
SWE-bench Verified    & 100 & 336 & 94.9\% \\
\bottomrule
\end{tabular}
\end{table}

The same checklist-synthesis procedure can be applied across SkillsBench, Terminal-Bench, and SWE-bench without changing its rules. On 20 unseen SkillsBench tasks, the frozen five-category requirement scheme covers 355 of 357 stated requirements, and all 214 generated criteria pass the mechanical validation checks. These results show that the checklist-synthesis procedure transfers to unseen tasks and benchmarks. We next examine whether the GCPC scores provide useful information about execution progress.

\paragraph{General-task diagnosis.}
On Terminal-Bench, GCPC assigns a higher score to the passing trajectory in
40 of 69 decidable same-task pairs, a lower score in only two pairs, and the
same score in 27 pairs (median $\Delta=+0.25$; $p<10^{-9}$). This suggests
that GCPC generally reflects differences between successful and unsuccessful
executions while providing finer-grained information than pass rate. For example, on \texttt{pypi-server}, the failing trajectory satisfies four
of six criteria: the package name, version, import, and function are correct,
but the agent does not build a distributable artifact or start the
package-index server. Pass rate records only that the task failed, whereas
GCPC shows that substantial progress was made and identifies the remaining
failures, assigning $0.67$ compared with $1.00$ for the passing trajectory.

\paragraph{Evidence boundary.}
SWE-bench reveals a limitation of log-only GCPC. Among 74 decidable
resolved/unresolved pairs, 58 receive the same GCPC score and 42.1\% of
criteria receive \abstain{}, indicating that many execution logs lack enough
evidence to distinguish successful from unsuccessful trajectories. For
example, on \texttt{sphinx-doc\_\_sphinx-10435}, both trajectories satisfy all
three criteria derived from the issue description and receive GCPC $1.00$;
only the official tests reveal a difference, after which the scores become
$1.00$ and $0.75$. GCPC therefore still applies to SWE-bench, but its
log-only informativeness is bounded by what the task instruction and
execution log make observable.



\section{Limitations}
\label{sec:limitations}
GCPC is limited by what the task instruction, verifier, and execution log make
observable. Provenance checks ensure that checklist criteria are traceable to
these sources, but do not guaranty that the sources capture every valid task
requirement. Similarly, when the execution log lacks sufficient evidence, GCPC
may provide little additional information, as shown by the 58 of 74 tied
SWE-bench pairs.

GCPC also depends on the judge and the official verifier. Criterion judgments
may vary across judge models, while GCPC with official outcome directly uses
the benchmark outcome and therefore inherits any verifier errors. Its stronger
agreement with PASS/FAIL should consequently not be interpreted as independent
validity evidence. In addition, our scalability claim refers to reducing manual checklist authoring across tasks, rather than reducing the computational cost of model inference.


\section{Conclusion}
We introduced GCPC, a verifier-grounded checklist evaluation framework that complements binary task outcomes with an auditable criterion-level view of execution progress. Rather than relying on either repeatedly authored human checklists or unconstrained model-generated criteria, GCPC combines reusable human-defined rules with task-specific checklist synthesis, provenance validation, evidence-based criterion judgment, and a separate deterministic official-outcome step. Experiments show that GCPC provides more informative trajectory evaluation than holistic scoring and reveals skill-associated improvements and regressions that remain hidden when the PASS/FAIL outcomes do not change. Reliability and transfer experiments further show that the same pipeline can be applied beyond skill-conditioned evaluation, while its informativeness remains limited by the evidence available in the trajectory.

\section*{Ethical Considerations}

The main negative societal impact we identify concerns misuse and harm that
can arise even when GCPC functions exactly as intended. Because GCPC assigns
partial credit to individually quotable, verifiable criteria rather than a
single opaque score, it is more attractive than a holistic judge as a
training reward signal for reinforcement learning: an agent could be
optimized directly against checklist wording. This is a genuine risk
precisely because the instrument works well at the level it was designed
for, criterion-level agreement with task requirements, and because that
same granularity gives an optimizer many small, quotable targets to
game rather than one coarse signal that is harder to locally exploit. An
agent could learn to produce log evidence that satisfies the letter of a
criterion, such as fabricated tool outputs or superficially matching phrasing,
without completing the underlying requirement, reproducing the Goodhart's
Law failure mode that motivated our evidence-grounded, abstention-aware
design in the first place, now at the level of the training signal rather
than the evaluation signal.

To mitigate this, we explicitly release the GCPC as an evaluation and audit
instrument, not as a validated reward function. Every design choice in this
paper (verbatim-quote provenance, verifier grounding, abstention on missing
evidence, and the reliability and enforceability checks of
\S\ref{sec:reliability}) targets measurement trustworthiness under a fixed
policy; none of it has been validated as a reward signal under active
optimization, where an agent can adapt its behavior specifically to exploit
the metric rather than merely being measured by it. We do not recommend using
GCPC to shape agent behavior via reinforcement learning without a separate
study of its robustness to such optimization pressure, analogous to the
gap between static benchmark performance and behavior under active
adversarial or reward-seeking pressure.

We do not identify a substantial concern about fairness, physical safety, or security specific to this contribution.

\bibliographystyle{ACM-Reference-Format}
\bibliography{references}

@inproceedings{merrill2026terminalbench,
  title     = {Terminal-Bench: Benchmarking Agents on Hard, Realistic Tasks
               in Command Line Interfaces},
  author    = {Merrill, Mike A. and Shaw, Alexander G. and Carlini, Nicholas
               and others},
  booktitle = {International Conference on Learning Representations},
  year      = {2026},
  note      = {arXiv:2601.11868}
}

@inproceedings{jimenez2024swebench,
  title     = {{SWE}-bench: Can Language Models Resolve Real-World
               {GitHub} Issues?},
  author    = {Jimenez, Carlos E. and Yang, John and Wettig, Alexander and
               Yao, Shunyu and Pei, Kexin and Press, Ofir and
               Narasimhan, Karthik R.},
  booktitle = {International Conference on Learning Representations},
  year      = {2024}
}

@inproceedings{qin2024infobench,
  author    = {Qin, Yiwei and Song, Kaiqiang and Hu, Yebowen and Yao, Wenlin and Cho, Sangwoo and Wang, Xiaoyang and Wu, Xuansheng and Liu, Fei and Liu, Pengfei and Yu, Dong},
  title     = {{InFoBench}: Evaluating Instruction Following Ability in Large Language Models},
  booktitle = {Findings of the Association for Computational Linguistics: ACL 2024},
  pages     = {13025--13048},
  year      = {2024},
  note      = {arXiv:2401.03601}
}

@misc{cook2024tick,
  author       = {Cook, Jonathan and Rockt{\"a}schel, Tim and Foerster, Jakob and Aumiller, Dennis and Wang, Alex},
  title        = {{TICK}ing All the Boxes: Generated Checklists Improve {LLM} Evaluation and Generation},
  year         = {2024},
  howpublished = {arXiv:2410.03608}
}

@inproceedings{lee2025checkeval,
  author    = {Lee, Yukyung and Kim, Joonghoon and Kim, Jaehee and Cho, Hyowon and Kang, Jaewook and Kang, Pilsung and Kim, Najoung},
  title     = {{CheckEval}: A Reliable {LLM}-as-a-Judge Framework for Evaluating Text Generation Using Checklists},
  booktitle = {Proceedings of the 2025 Conference on Empirical Methods in Natural Language Processing},
  year      = {2025},
  note      = {arXiv:2403.18771}
}

@inproceedings{rlcf2025,
  author    = {Viswanathan, Vijay and Sun, Yanchao and Ma, Shuang and Kong, Xiang and Cao, Meng and Neubig, Graham and Wu, Tongshuang},
  title     = {Checklists Are Better Than Reward Models For Aligning Language Models},
  booktitle = {Advances in Neural Information Processing Systems (NeurIPS)},
  year      = {2025},
  note      = {arXiv:2507.18624}
}

@misc{liu2026openrubrics,
  author       = {Liu, Tianci and Xu, Ran and Yu, Tony and Hong, Ilgee and Yang, Carl and Zhao, Tuo and Wang, Haoyu},
  title        = {{OpenRubrics}: Towards Scalable Synthetic Rubric Generation for Reward Modeling and {LLM} Alignment},
  year         = {2025},
  howpublished = {arXiv:2510.07743}
}

@misc{lin2024wildbench,
  author       = {Lin, Bill Yuchen and Deng, Yuntian and Chandu, Khyathi and Brahman, Faeze and Ravichander, Abhilasha and Pyatkin, Valentina and Dziri, Nouha and Le Bras, Ronan and Choi, Yejin},
  title        = {{WildBench}: Benchmarking {LLM}s with Challenging Tasks from Real Users in the Wild},
  year         = {2024},
  howpublished = {arXiv:2406.04770}
}

@misc{interacteval2024,
  author       = {Chu, SeongYeub and Kim, JongWoo and Yi, MunYong},
  title        = {Think Together and Work Better: Combining Humans' and {LLM}s' Think-Aloud Outcomes for Effective Text Evaluation},
  year         = {2024},
  howpublished = {arXiv:2409.07355}
}

@inproceedings{dineen2025qalign,
  author    = {Dineen, Jacob and RRV, Aswin and Liu, Qin and Xu, Zhikun and Ye, Xiao and Shen, Ming and Li, Zhaonan and Lu, Shijie and Baral, Chitta and Chen, Muhao and Zhou, Ben},
  title     = {{QA-LIGN}: Aligning {LLM}s through Constitutionally Decomposed {QA}},
  booktitle = {Findings of the Association for Computational Linguistics: EMNLP 2025},
  pages     = {20619--20642},
  year      = {2025},
  note      = {arXiv:2506.08123}
}

@misc{healthbench2025,
  author       = {Arora, Rahul K. and others},
  title        = {{HealthBench}: Evaluating Large Language Models Towards Improved Human Health},
  year         = {2025},
  howpublished = {arXiv:2505.08775}
}

@misc{que2024hellobench,
  author       = {Que, Haoran and Duan, Feiyu and He, Liqun and Mou, Yutao and Zhou, Wangchunshu and Liu, Jiaheng and Rong, Wenge and Wang, Zekun Moore and Yang, Jian and Zhang, Ge and Peng, Junran and Zhang, Zhaoxiang and Zhang, Songyang and Chen, Kai},
  title        = {{HelloBench}: Evaluating Long Text Generation Capabilities of Large Language Models},
  year         = {2024},
  howpublished = {arXiv:2409.16191}
}

@inproceedings{wei2025rocketeval,
  title={Rocketeval: Efficient automated LLM evaluation via grading checklist},
  author={Wei, Tianjun and Wen, Wei and Qiao, Ruizhi and Sun, Xing and Ma, Jianghong},
  booktitle={International Conference on Learning Representations},
  volume={2025},
  pages={58593--58619},
  year={2025}
}

@inproceedings{zhou2026autochecklist,
  title={AutoChecklist: Composable Pipelines for Checklist Generation and Scoring with LLM-as-a-Judge},
  author={Zhou, Karen and Tan, Chenhao},
  booktitle={Proceedings of the 64th Annual Meeting of the Association for Computational Linguistics (Volume 3: System Demonstrations)},
  pages={515--525},
  year={2026}
}

@inproceedings{song2024finesure,
  author    = {Song, Hwanjun and Su, Hang and Shalyminov, Igor and Cai, Jason and Mansour, Saab},
  title     = {{FineSurE}: Fine-grained Summarization Evaluation using {LLM}s},
  booktitle = {Proceedings of the 62nd Annual Meeting of the Association for Computational Linguistics},
  year      = {2024}
}

@inproceedings{saadfalcon2025unittests,
  author    = {Saad-Falcon, Jon and Vivek, Rajan and Berrios, William and Naik, Nandita Shankar and Franklin, Matija and Vidgen, Bertie and Singh, Amanpreet and Kiela, Douwe and Mehri, Shikib},
  title     = {{LMUnit}: Fine-grained Evaluation with Natural Language Unit Tests},
  booktitle = {Findings of the Association for Computational Linguistics: EMNLP 2025},
  pages     = {3303--3324},
  year      = {2025},
  note      = {arXiv:2412.13091}
}

@misc{zhang2026rubricbench,
  author       = {Zhang, Qiyuan and Zhou, Junyi and Wang, Yufei and Lyu, Fuyuan and Ming, Yidong and Xu, Can and Sun, Qingfeng and Zheng, Kai and Kang, Peng and Liu, Xue and Ma, Chen},
  title        = {{RubricBench}: Aligning Model-Generated Rubrics with Human Standards},
  year         = {2026},
  howpublished = {arXiv:2603.01562}
}

@inproceedings{furuhashi2025checklists,
  author    = {Furuhashi, Momoka and Nakayama, Kouta and Kodama, Takashi and Sugawara, Saku},
  title     = {Are Checklists Really Useful for Automatic Evaluation of Generative Tasks?},
  booktitle = {Proceedings of the 2025 Conference on Empirical Methods in Natural Language Processing},
  year      = {2025},
  note      = {arXiv:2508.15218}
}

@misc{ye2024justice,
  author       = {Ye, Jiayi and Wang, Yanbo and Huang, Yue and Chen, Dongping and Zhang, Qihui and Moniz, Nuno and Gao, Tian and Geyer, Werner and Huang, Chao and Chen, Pin-Yu and Chawla, Nitesh V. and Zhang, Xiangliang},
  title        = {Justice or Prejudice? Quantifying Biases in {LLM}-as-a-Judge},
  year         = {2024},
  howpublished = {arXiv:2410.02736}
}

@inproceedings{siro2026learning,
  author    = {Siro, Clemencia and Aliannejadi, Pourya and Aliannejadi, Mohammad},
  title     = {Learning to Judge: {LLM}s Designing and Applying Evaluation Rubrics},
  booktitle = {Findings of the Association for Computational Linguistics: EACL 2026},
  pages     = {6371--6389},
  year      = {2026}
}

@inproceedings{ma2024agentboard,
  author    = {Ma, Chang and Zhang, Junlei and Zhu, Zhihao and Yang, Cheng and Yang, Yujiu and Jin, Yaohui and Lan, Zhenzhong and Kong, Lingpeng and He, Junxian},
  title     = {{AgentBoard}: An Analytical Evaluation Board of Multi-turn {LLM} Agents},
  booktitle = {Advances in Neural Information Processing Systems (NeurIPS), Datasets and Benchmarks Track},
  year      = {2024},
  note      = {Oral. arXiv:2401.13178}
}

@misc{pan2024webcanvas,
  author       = {Pan, Yichen and Kong, Dehan and Zhou, Sida and Cui, Cheng and Leng, Yifei and Jiang, Bing and Liu, Hangyu and Shang, Yanyi and Zhou, Shuyan and Wu, Tongshuang and others},
  title        = {{WebCanvas}: Benchmarking Web Agents in Online Environments},
  year         = {2024},
  howpublished = {arXiv:2406.12373}
}

@inproceedings{yao2025taubench,
  author    = {Yao, Shunyu and Shinn, Noah and Razavi, Pedram and Narasimhan, Karthik},
  title     = {$\tau$-bench: A Benchmark for Tool-Agent-User Interaction in Real-World Domains},
  booktitle = {International Conference on Learning Representations (ICLR)},
  year      = {2025},
  note      = {arXiv:2406.12045}
}

@inproceedings{lightman2024verify,
  author    = {Lightman, Hunter and Kosaraju, Vineet and Burda, Yura and Edwards, Harri and Baker, Bowen and Lee, Teddy and Leike, Jan and Schulman, John and Sutskever, Ilya and Cobbe, Karl},
  title     = {Let's Verify Step by Step},
  booktitle = {International Conference on Learning Representations (ICLR)},
  year      = {2024},
  note      = {arXiv:2305.20050}
}

@misc{agentprocessbench2026,
  author       = {Fan, Shengda and Ye, Xuyan and Huo, Yupeng and Chen, Zhi-Yuan and Guo, Yiju and Yang, Shenzhi and Yang, Wenkai and Ye, Shuqi and Chen, Jingwen and Chen, Haotian and Cong, Xin and Lin, Yankai},
  title        = {{AgentProcessBench}: Diagnosing Step-Level Process Quality in Tool-Using Agents},
  year         = {2026},
  howpublished = {arXiv:2603.14465}
}

@inproceedings{sharma2026willful,
  author    = {Sharma, Reshabh K. and Barke, Shraddha and Zorn, Benjamin},
  title     = {Willful Disobedience: Automatically Detecting Failures in Agentic Traces},
  booktitle = {Proceedings of the ACM Conference on AI and Agentic Systems (CAIS)},
  year      = {2026},
  doi       = {10.1145/3786335.3813153},
  note      = {arXiv:2603.23806}
}

@misc{odysseys2026,
  author       = {Jang, Lawrence Keunho and Koh, Jing Yu and Fried, Daniel and Salakhutdinov, Ruslan},
  title        = {Odysseys: Benchmarking Web Agents on Realistic Long Horizon Tasks},
  year         = {2026},
  howpublished = {arXiv:2604.24964}
}

@inproceedings{gou2025mind2web2,
  author    = {Gou, Boyu and Huang, Zanming and Ning, Yuting and Gu, Yu and Lin, Michael and Qi, Weijian and Kopanev, Andrei and Yu, Botao and Jim{\'e}nez Guti{\'e}rrez, Bernal and Shu, Yiheng and Song, Chan Hee and Wu, Jiaman and Chen, Shijie and Moussa, Hanane Nour and Zhang, Tianshu and Xie, Jian and Li, Yifei and Xue, Tianci and Liao, Zeyi and Zhang, Kai and Zheng, Boyuan and Cai, Zhaowei and Rozgic, Viktor and Ziyadi, Morteza and Sun, Huan and Su, Yu},
  title     = {{Mind2Web 2}: Evaluating Agentic Search with Agent-as-a-Judge},
  booktitle = {Advances in Neural Information Processing Systems (NeurIPS), Datasets and Benchmarks Track},
  year      = {2025},
  note      = {arXiv:2506.21506}
}

@misc{ding2026adarubric,
  author       = {Ding, Liang},
  title        = {{AdaRubric}: Task-Adaptive Rubrics for Reliable {LLM} Agent Evaluation and Reward Learning},
  year         = {2026},
  howpublished = {arXiv:2603.21362},
  note         = {KnowFM Workshop @ ACL 2026}
}

@inproceedings{wan2026rubricguided,
  author    = {Wan, Yuxuan and Fang, Tianqing and Li, Zaitang and Huo, Yintong and Wang, Wenxuan and Mi, Haitao and Yu, Dong and Lyu, Michael R.},
  title     = {Inference-Time Scaling of Verification: Self-Evolving Deep Research Agents via Test-Time Rubric-Guided Verification},
  booktitle = {Findings of the Association for Computational Linguistics: ACL 2026},
  year      = {2026},
  note      = {arXiv:2601.15808}
}

@inproceedings{zheng2023mtbench,
  author    = {Zheng, Lianmin and Chiang, Wei-Lin and Sheng, Ying and Zhuang, Siyuan and Wu, Zhanghao and Zhuang, Yonghao and Lin, Zi and Li, Zhuohan and Li, Dacheng and Xing, Eric P. and Zhang, Hao and Gonzalez, Joseph E. and Stoica, Ion},
  title     = {Judging {LLM}-as-a-Judge with {MT}-Bench and Chatbot Arena},
  booktitle = {Advances in Neural Information Processing Systems (NeurIPS), Datasets and Benchmarks Track},
  year      = {2023},
  note      = {arXiv:2306.05685}
}

@article{wang2023voyager,
  author  = {Wang, Guanzhi and Xie, Yuqi and Jiang, Yunfan and Mandlekar, Ajay and Xiao, Chaowei and Zhu, Yuke and Fan, Linxi and Anandkumar, Anima},
  title   = {Voyager: An Open-Ended Embodied Agent with Large Language Models},
  journal = {Transactions on Machine Learning Research},
  year    = {2024},
  note    = {arXiv:2305.16291}
}

@inproceedings{xu2026skills,
  author    = {Xu, Renjun and Yan, Yang},
  title     = {Agent Skills for Large Language Models: Architecture, Acquisition, Security, and the Path Forward},
  booktitle = {Agent Skills '26 Workshop @ ACM Conference on AI and Agentic Systems (CAIS)},
  year      = {2026},
  note      = {arXiv:2602.12430}
}

@misc{li2026skillsbench,
  author       = {Li, Xiangyi and Liu, Yimin and Chen, Wenbo and You, Bingran and Di, Zonglin and He, Yifeng and others},
  title        = {{SkillsBench}: Benchmarking How Well Agent Skills Work Across Diverse Tasks},
  year         = {2026},
  howpublished = {arXiv:2602.12670}
}

@misc{skilljuror2026,
  author       = {Chen, Zhiyu and Guo, Zihan and Huang, Bo and Lu, Bingwei and Lin, Jianghao and Zhou, Yuanjian and Zhang, Weinan},
  title        = {{SkillJuror}: Measuring How Agent Skill Organization Changes Runtime Behavior},
  year         = {2026},
  howpublished = {arXiv:2606.11543}
}

@misc{skillsbench2026granularity,
  author       = {Xu, Xiaonan and Wu, Wenjing},
  title        = {Skill Availability and Presentation Granularity in Large-Language-Model Agents: A Controlled {SkillsBench} Study},
  year         = {2026},
  howpublished = {arXiv:2605.31408}
}

@misc{skillsurvey2026,
  author       = {Ding, Kexin and Zhou, Yang and Jin, Can and Tong, Feng and Zhou, Mu and Metaxas, Dimitris N.},
  title        = {Agent Skill Evaluation and Evolution: Frameworks and Benchmarks},
  year         = {2026},
  howpublished = {arXiv:2606.11435}
}

@misc{jiang2026sok,
  author       = {Jiang, Yanna and Li, Delong and Deng, Haiyu and Ma, Baihe and Wang, Xu and Wang, Qin and Yu, Guangsheng},
  title        = {SoK: Agentic Skills -- Beyond Tool Use in LLM Agents},
  year         = {2026},
  howpublished = {arXiv:2602.20867}
}

@misc{li2026ecosystem,
  author       = {Li, Hao and Mu, Chunjiang and Chen, Jianhao and Ren, Siyue and Cui, Zhiyao and Zhang, Yiqun and Bai, Lei and Hu, Shuyue},
  title        = {Organizing, Orchestrating, and Benchmarking Agent Skills at Ecosystem Scale},
  year         = {2026},
  howpublished = {arXiv:2603.02176}
}

@inproceedings{kapoor2026holistic,
  title={Holistic Agent Leaderboard: The Missing Infrastructure for {AI} Agent Evaluation},
  author={Kapoor, Sayash and Stroebl, Benedikt and Kirgis, Peter and Nadgir, Nitya and Siegel, Zachary and Wei, Boyi and Xue, Tianci and Chen, Ziru and Chen, Felix and Utpala, Saiteja and others},
  booktitle={International Conference on Learning Representations},
  volume={2026},
  pages={98778--98849},
  year={2026}
}

@inproceedings{mohammadi2025evaluation,
  title={Evaluation and Benchmarking of {LLM} Agents: A Survey},
  author={Mohammadi, Mahmoud and Li, Yipeng and Lo, Jane and Yip, Wendy},
  booktitle={Proceedings of the 31st ACM SIGKDD Conference on Knowledge Discovery and Data Mining V. 2},
  pages={6129--6139},
  year={2025}
}

\end{document}